# Bonding Nature in MgB$_2$

E.Nishibori, M.Takata*, M.Sakata, H.Tanaka[1], T.Muranaka[2], J.Akimitsu[2]

*Department of Applied Physics, Nagoya University, Nagoya 464-8603, Japan*
[1]*Department of Material Science, Shimane University, Matsue 690-8504, Japan*
[2]*Department of Physics, Aoyama-Gakuin University, Chitosedai, Setagaya-ku, Tokyo 157-8572, Japan*

**abstract**
The accurate charge density of MgB$_2$ was observed at room temperature(R.T.) and 15K by the MEM(Maximum Entropy Method)/Rietveld analysis using synchrotron radiation powder data. The obtained charge density clearly revealed the covalent bonding feature of boron forming the 2D honeycomb network in the basal plane, on the other hand, Mg is found to be in divalent state. A subtle but clear charge concentration was found on boron 2D sheets at 15K, which should be relating to superconductivity.

KEYWORDS: MgB$_2$, Superconductivity, Bonding Nature, Charge Density, Synchrotron Radiation, Maximum Entropy Method

* a41024a@nucc.cc.nagoya-u.ac.jp

Since the recent discovery of the superconductivity in MgB$_2$[1], the anomalous number of experimental and theoretical studies has been performed to understand the mechanism of superconductivity at comparatively high Tc, 39K. The crystal structure of this material has been known as hexagonal (AlB$_2$ type, space group *P6/mmm*) since 1954[2]. The characteristic boron honeycomb sheets are sandwiched between the Mg triangular sheets like an intercalated graphite. The band structure calculations[3-6] have predicted the existence of charge donation of two electrons from the ionized Mg to the boron conduction band while strong B-B covalent bonding is retained. Then, the superconductivity in MgB$_2$, which should essentially due to the metallic nature of the Boron 2D sheets, has been interpreted as the phonon-mediated BCS-type mechanism. Such a two dimensional structure is the common feature to the oxide superconductors as well as the intercalated graphite. Very recently, the doping on the Mg site or Boron 2D sheets has been carried out to reveal the effect of the electron concentration on the superconducting temperature[7-10]. Several reports for the loss of superconductivity have been presented for Mg$_{1-x}$Al$_x$B$_2$[7], MgB$_{1-x}$C$_x$[8], Mg$_{1-x}$Li$_x$B$_2$[9] and Mg$_{1-x}$Mn$_x$B$_2$[10]. The pressure evolution of superconducting transition temperature was also reported and discussed for the relation with the B-B and Mg-B bonding distances[11,12]. In the mean time, the structural information used in the discussions has been limited within atomic level, i.e. lattice constants, bonding distance, etc. An experimental charge density would give a better understanding how the superconductivity links to the electronic and crystal structure of $MgB_2$. Here, we report accurate charge densities of $MgB_2$ at R.T. and 15K using synchrotron radiation powder data and present the experimental evidences for the strong B-B covalent bonding, full ionization of Mg atoms at both temperature as well as the charge concentration on boron 2D sheet at 15K, which should be relating to superconducting mechanism.

The $MgB_2$ sample used in this work was prepared by heating a pressed pellet of stoichiometric amounts of Mg and amorphous B for 10 hours at 700°C under an argon pressure of 196 MPa and was superconducting with $T_c$= 39 K. The granularity of the powder was made even less than 3-micron diameter by the precipitation method in order to get a homogeneous intensity distribution in Debye-Scherrer powder ring. The obtained powder sample was sealed in a silica glass capillary (0.3mm int. diam.). The synchrotron radiation x-ray powder experiment with imaging plate (IP) as detectors was carried out by Large Debye-Scheerer Camera at SPring-8 BL02B2 . The He gas circulation type cryostat was used for the measurement at low temperature. The x-ray powder patterns were measured at room temperature (R.T.) and 15K($<<T_c$). The both data were obtained under the same experimental conditions except for the temperature. The exposure time was 1hours. The wavelength of incident x-rays was 0.6 . The x-ray powder pattern of $MgB_2$ was obtained with a 0.02°step from 9.0° to 65.0° in 2 , which corresponds to 0.57 resolution in $d$-spacing.

The charge density distributions at both temperatures were visualized by the MEM/Rietveld method, which is a combination of the MEM and the Rietveld refinement[13,14]. The method has been successfully applied to the charge density studies of fullerene compounds[13,15-17], intermetallic compounds[18], α-boron[19], manganite[20], etc. For instance, the MEM/Rietveld method using synchrotron radiation powder data has revealed Mn $3d_{x^2-y^2}$ orbital order as a Mn-O bonding electron distribution associated with Mn(3$d$)-O(2$p\sigma$) hybridization at antiferromagnetic state in manganite, $NdSr_2Mn_2O_7$ [20]. The details of the method are described in the previous papers[13-17]. In the present powder data, several weak impurity peaks were found and identified as MgO. The impurity MgO phase was also taken into account in the Rietveld pre-analysis. The space group was assigned to be *P6/mmm* for the both data at R.T. and 15K. This implies that there is no structural phase transition from R.T. to 15K. The results of the Rietveld refinement are shown for R.T. and 15K in Fig. 1(a) and 1(b), respectively. The refined lattice parameters were listed in Table 1. The weighted profile reliability factors of the Rietveld refinement as a pre-analysis for the MEM, $R_{WP}$, were 4.7% and 2.6% for R.T. and 15K, respectively. And the reliability factors based on the integrated intensities, $R_I$, were 3.1% and 3.4% for R.T. and 15K, respectively. In the analysis, the observed structure factors of the 55 reflections were derived from the observed integrated intensities. Then, they were used for the further MEM analysis. Following to the Rietveld pre-analysis, the MEM analysis was carried out using 64× 64× 72 pixels. The reliable factors of the final MEM charge density were 1.7% and 1.5% for R.T. and 15K, respectively.

A three-dimensional representation of the final MEM charge density at 15K is shown in Fig.2 as an equi-charge density surface with structure model. The equi-density level is 0.75 e/Å$^3$. The obtained MEM charge density clearly exhibits the strong covalent bonding network of boron 2D sheet forming the six-membered rings, which are colored in blue. On the other hand, there are no localized electron densities between Mg and boron atoms. In the interatomic region, electrons are distributed rather evenly similar to metal bonding. This characteristic density features are preserved in the charge density obtained at R.T and consistent with the calculated band structures indicating two bands model[3]. Based on this model, several theoretical mechanisms of superconductivity have been proposed[21,22].

In Fig. 3, the MEM charge densities of the (110) sections containing Mg and boron atoms are shown for R.T. and 15K. The contour lines are drawn only for the lower density region. It is confirmed that there is no significant overlapping of the charge density around Mg atomic sites. This is a high contrast to that of the boron-boron network. The MEM charge densities clearly reveal the boron-boron covalent bonding features. Although the change of the boron-boron interatomic distance is extremely small between R.T. and 15K as shown in Table 1, the charge density values at the bond midpoints show the distinct different values, which are 0.9 and 1.0 eÅ$^{-3}$ at R.T. and 15K, respectively. These values are in the range between those of Si (0.7eÅ$^{-3}$)[14] and Diamond(1.4eÅ$^{-3}$)[14] and very close to the value of hexagonal-BN(1.0eÅ$^{-3}$)[23].

The valence of the atom was examined by accumulating the number of electrons around a certain atom in the MEM density. So far, the valence of metal atoms encapsulated in metallofullerene has been determined experimentally from the MEM charge densities[13-17]. The number of electrons around Mg atom was estimated as about 10.0(1)$e$ and 10.0(1)$e$, respectively. These values are very close to the number of electrons of Mg$^{2+}$ ion. This means that the Mg atoms are fully ionized in MgB$_2$ crystal at both R.T. and 15K. On the other hand, the number of electrons belonging to the boron 2D sheets shows significant difference, they are 9.9(1)$e$ and 10.9(1)$e$ at R.T. and 15K, respectively. This can be interpreted that the valence of the whole boron 2D sheet changes from neutral to monovalent, i.e., (B-B)$^-$ at 15K, which evidences the increase of charge at the B-B bond midpoint at 15K. Though the full charge transfer from Mg to boron 2D sheets had been expected to occur and forming isoelectronic sheet with graphite, a simple direct full charge transfer from Mg$^{2+}$ to boron 2D sheet was not observed. The present results support the following scenario, that is, the valence electrons are delocalized in the inter-atomic region at R.T., and half of them localized on the boron 2D sheets at low temperature. This scenario implies the presence of the electron transfer from π bonds consisting of $p_z$ orbitals to in-plain σ bonds consisting of $p_{xy}$ orbitals in the two bands model of MgB$_2$ at 15K.

In conclusion, the strong covalent boron network and the full ionization of Mg were found in $MgB_2$ charge density at both R.T. and 15K by the MEM/Rietveld method using synchrotron radiation powder data. In addition, a subtle but important charge concentration on boron 2D sheet at 15K was also found, which should be relating to the superconductivity of this compound. Our finding on the bonding nature of $MgB_2$ would become useful information for a better understanding on the superconductivity of $MgB_2$ and/or exploring $MgB_2$ relating superconductor.

This work has been partially supported by a Grant-in-Aid for Science Research from the Ministry of Education, Science, Sports and Culture, Japan. This work was also supported by the Toyota-Riken, the Murata Science Foundation and the Sumitomo Foundation. The synchrotron radiation experiments were performed at SPring-8 BL02B2 with the approval of the Japan Synchrotron Radiation Research Institute (JASRI).

**Table 1.** The lattice parameters and B-B, Mg-B atomic distances determined by the Rietveld analysis for MgB$_2$ at R.T. and 15K.

| Parameter | R.T.(Å) | 15K(Å) |
|---|---|---|
| Lattice Parameters | a = 3.08831(3) <br> c = 3.52415(8) | a = 3.08365(2) <br> c = 3.51504(4) |
| B-B distance | 1.78304(1) | 1.78035(1) |
| Mg-B distance | 2.50682(6) | 2.50170(3) |

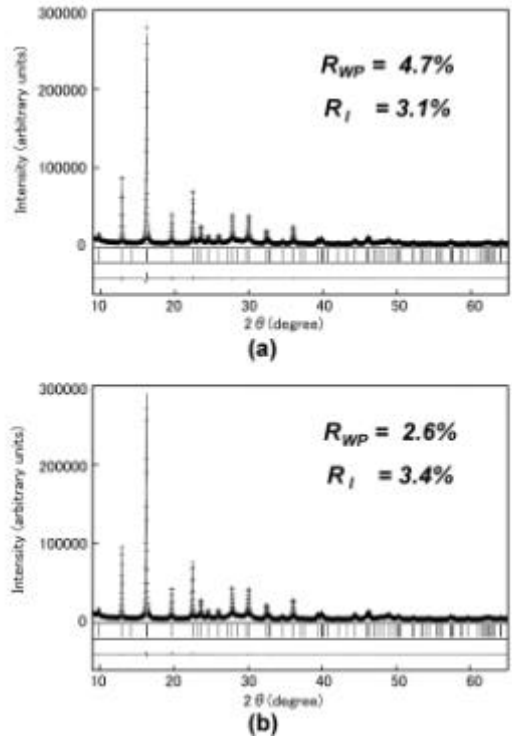

Fig.1 Fitting results of Rietveld analysis of MgB$_2$ at (a)R.T. and (b) 15K.

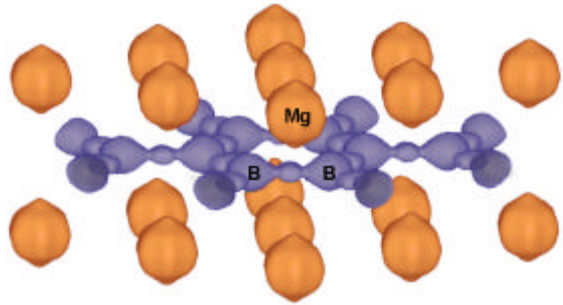

**Fig.2** The equi-contour(0.75 e/Å$^3$) surface of MEM charge density of MgB$_2$ at 15K.

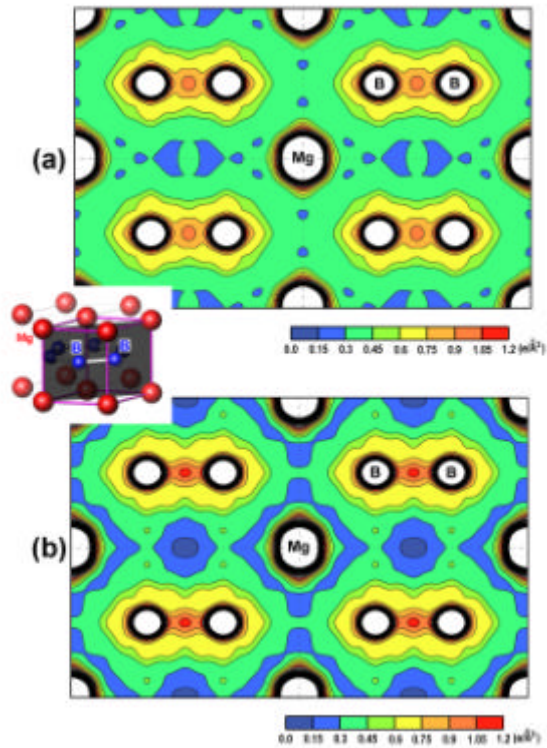

**Fig.3** The (110) sections of the MEM Charge Density of MgB$_2$ at (a)R.T. and (b)15K with the schematic representation of the crystal structure. The contour lines are drawn from 0.0 to 3.9 at 0.15(e/Å$^3$) intervals for four unit cells.